# FLOW IMPROVING DEVICES FOR MARINE PROPULSERS.
# CONCEPT DESIGN STUDY.

**Vladimir Sluchak**

## A. Concept description

*Flow-improving devices (FIDs) are used the most effectively for the reduction of noise and vibration induced by a propulsor. These devices may also increase the efficiency of a propulsor and in some cases should be considered and designed as a part of an Integrated Propulsor, for example, non-axis symmetrical ducts.*

*The FIDs may be naturally subdivided into two types: active and passive. The active ones are powered, such as the devices for local suction and injection of water, air, or polymer solutions, devices for local heating, etc. Passive FIDs are the ones that, generally speaking, use the power of inflow. Sometimes the power is lost unless not used for flow improvement – this is how the energy saving is achieved. Many passive FIDs are based on the effect of longitudinal vortices working in a boundary layer of a hull. The traversal velocities induced by these vortices redistribute the longitudinal velocities in a boundary layer of a hull and thus affect the circumferential non-uniformity of a propeller's inflow.*

*The vortices can be produced differently. The most often used solution is a simple flat wing of a small aspect ratio placed normally to the surface of a hull forward of the propeller in the stern part of the hull, usually merged into a thick boundary layer. These wings affect the flow downstream with their trailing vortices.*

*Some of the FIDs redirect the flow with their head vortices, for example, the ones installed on Russian SSBN Akula (Typhoon). These were shaped specially to produce smooth local 3-dimensional separation of boundary layer inducing a pair of head vortices of the same power and opposite rotation.*

*Apart from these vortex-generating devices are arrangements with special profiling of the front edge of stabilizers. These are aimed at eliminating the head vortices of the appendages right at the place of their origination. The pre–swirls, pre-ducts, and nonsymmetrical ducts are the energy-saving FIDs usually placed close to the propeller. These affect the propeller's inflow directly - via the pressure field.*

The core of several new solutions of FIDs proposed in this document is a non-axis-symmetrical ring wing placed upstream of the propeller between the propeller and the appendages.
Several arrangements are considered in which the Ring Wing functions differently but always delivers a positive effect on the Propulsor.
a) The base arrangement in which the thin and narrow Ring Wing is placed relatively far forward the Propeller (usually as far as the first met appendages) and sometimes is fixed on or built into the appendages or stabilizers of the Marine Vehicle. This is the case when the ring affects the flow with its longitudinal vortices (Fig 1,2 and theory in section B).
b) The arrangements where a Ring Wing works mostly as an effective stabilizer substituting the conventional cross-shaped one (Fig 2).
c)The arrangements in which a non-axis-symmetrical Ring Wing is placed close to the propeller and works as a pre-duct or full duct so the Ring and the Propeller should be considered as an unconventional Integrated Propulsor.
Arrangements 1, 2, and 5 of Fig. 2 have been patented by the author.
Section G of this document describes the interactive design procedure for these devices. This procedure is novel and may be used for any device and architectural element aimed at reducing the noise and vibration of a propeller (especially discrete sound).

## B. Theoretical Basis

The currently admitted model of the flow filtration by a propeller allows the formulation of the following flow improvement targets:

-to decrease a high frequency (with continuous spectrum) component of the propeller noise the rough characteristic of the flow non-uniformity forwards the propeller is to be decreased - the difference $\Delta$ between maximum and minimum of the circular distribution **V($\varphi$,r)** of the velocity at end sections of blades (Fig. 3).

- to decrease discrete spectrum components of the vibration and the discrete spectrum sound or "revolution sound" of the propeller it is enough to decrease amplitudes and sums of amplitudes $|\mathbf{a}_{qz}|$ ; $|\mathbf{a}_{qz\pm1}|$ ; $|\mathbf{b}_{qz}|$ ; $|\mathbf{b}_{qz\pm1}|$ of harmonics ( Fourier series members) of **V($\varphi$,r)** at the most charged sections of propeller blades. Here: **Z** is the number of propeller blades ; **q** - digit ; **a , b** - cosine and sine harmonic amplitudes.

The harmonics with numbers more than **Z** are very hard to control because they practically never dominate in **V($\varphi$, r)** making hundredths of $\Delta$. So the second part of the flow improvement problem is much more delicate than the first one and actually has never been approached effectively enough. Usually, the discrete sound of the propeller is optimized by some expensive way of propeller design which in particular implies an increase of the number of blades. The described flow improvement problem is principally solved in two ways: the optimization of the vessel architecture (the shape and dimensions of hull, sail, stabilizers, covers, etc. ) and the usage of special flow improvement devices (FIDs).

For different kinds of Marine Vehicles (submarines and submersibles - at most), a ring-shaped wing placed forward of the propeller around the tale of a hull presents a very effective combination of these two ways of flow improvement. It reduces the non-uniformity of the flow directly and indirectly: being an effective vortex-generating FID and permitting a decrease of stabilizer dimensions.

One may consider the following architecture for a submarine with a pushing propulsor as an ideal: axis-symmetrical hull, no sail, stabilizers as a ring-shaped wing on relatively narrow struts or pillars. All the real architectural types might be put in some order of deviation from it.

The more architecture deviates from the ideal one (for instance - the larger the sail would be ) the more flow-improving work would be required. Keeping the ring-shaped wing within the architecture allows to do this work effectively.

All the arrangements with a Ring Wing presented in Fig 2 lead to flow improvement and have some additional advantages. These constructions must be subdivided into two categories:
1) with a ring wing working as a flow-improving device affecting the downstream flow by the longitudinal trailing vortices (this type will be further referred to as RFID – Ring-shaped Flow Improving Device) and
2) with the ring wing working mostly as a stabilizer.

In the first type of arrangement, the ring wing affects the flow by its passively generated longitudinal vortices. The intensity of an elementary vortex running down from the ring is proportional to the derivative
$d(\mathbf{V} \cdot \mathbf{b} \cdot \varepsilon) / d\varphi$ , where **V** is the velocity of the flow on the front of the ring, **b** - the chord of the ring, $\varepsilon$ - local angle of attack composed of a constructive angle of the ring wing and the angle between the flow direction and the axis of the ring. As you can see from that formula the vortex shroud running down from a ring wing is sensitive to the irregularity of the flow upstream: the more intense the irregularity the greater the intensity of generated vortices. The vortices intensity may be controlled precisely at any sector of the propeller's disk with the variation of the ring geometry which the simplest case is a circular variation of the chord. The generated vortices work aft of the ring in the boundary layer of a hull principally the same way as for most other passive vortex-generating FIDs: they induce traversal flows that redistribute the layers with different longitudinal velocities. The details of the Ring Wing Flow Improving Device theory may be found in [1]. The theory of harmonics control used for the iterative design or geometry tuning of an RFID is described in section G of this document.

Partial or total substitution of cross-shaped stabilizers with a ring wing in most of the arrangements of Fig.2 delivers a decrease of flow non-uniformity indirectly: the hydrodynamic wake of the initial cross-shaped stabilizer making directly into the area swept by the propeller gets substituted with a wake of relatively narrow struts of the ring wing. The number of struts and their location can be optimized. The substitution of a cross-shaped wing with a ring wing can also deliver the gain in the stabilizer's span. In some cases, the span may be set even lower than the maximum diameter of a hull, which is important for small portable submarines and submersibles. The integrated comparison of a cross-shaped wing to a ring wing proves this point (Table 1 below and Fig 4).

Table1

| Wing's type  Coefficients | Flat wing of a finite span  − | Cross-shaped wing composed of two equal parts  + | Ring - shaped wing  o |
|---|---|---|---|
| Derivative of a coefficient of a lifting force $C_y^\alpha$ | $\dfrac{2\pi}{1+\dfrac{2}{\lambda_-}}$ | $\dfrac{\pi}{1+\dfrac{2}{\lambda_+}}$ | $\dfrac{\pi}{1+\dfrac{\pi}{2\lambda_O}}$ |
| Coefficient of Induced Drag $C_X^I$ | $\dfrac{C_{y-}^2}{\pi\lambda_-}$ | $\dfrac{2C_{y+}^2}{\pi\lambda_+}$ | $\dfrac{C_{yO}^2}{2\lambda_O}$ |
| Coefficient of Induced Drag adjusted (without sucking force) $C_X^{I'}$ | $C_y^{\alpha_0} + \dfrac{C_{y-}^2}{C_{y-}^\alpha}$ | $C_y^{\alpha_0} + \dfrac{C_{y+}^2}{C_{y+}^\alpha}$ | $2\varepsilon_O^2\pi + \dfrac{C_{yO}^2}{C_{yO}^\alpha}$ |

Designations in Table1 and Fig 4 :

$\lambda$            aspect ratio of a wing,

$\alpha_O$          angle between the chord and line of zero lifting force of a Flat Wing,

$\varepsilon_O$          the opening angle of a ring wing,

l , b          span and chord of a Flat or Cross Wing ,

D,b          diameter (span) and chord of a Ring Wing ,

$\lambda_+ = \dfrac{l}{b}; \lambda_O = \dfrac{D}{b}$      aspect ratios of a cross wing and a ring wing accordingly;

S          one side area of the wing;

$S_\Pi = \dfrac{C_y^\alpha}{2\pi} S$      the effective area of the wing;

$S_I = C_X^I S$      "inductive" area of the wing;

The ratios between areas of Ring Wing and Cross Wing:

$$\overline{S}^O = \dfrac{S^O}{S^+} \;;\; \overline{S}_\Pi^O = \dfrac{S_\Pi^O}{S_\Pi^+} \;;\; \overline{S}_I^O = \dfrac{S_I^O}{S_I^+}$$

$\overline{l} = \dfrac{l}{D}$          spans' ratio

The ratio of effective areas is equal to the ratio of lifting forces at the same inflow velocity and angles of attack, and the ratio of "inductive" areas is equal to the ratio of induced drag under the same conditions.
As one can see from the diagrams of Fig.4 for example:

for $\lambda_+ < 4$ and given constant lifting force, ~20% gain in stabilizers span may be received with simultaneous gain in area and inductive drag.

## C. Expected Improvement.

*The expected improvement depends upon the type of vessel. The results of the implementation of the proposed arrangements are expected to be similar to the ones received on the models tested in the experimental basin of Krylov State Research Center. These results are presented in Fig.3,5,6,7,8. The model tests with RFID have been carried out on large models ( about 20 feet long) in a big towing tank of Krylov State Research Center. The calculations of the discrete (revolution) sound were performed using the original method, and a computer program developed at the same research center (Bavin et al.). The inputs for the program are the field of flow velocities $V(\varphi,r)$ forward of the propeller, the propeller's detailed geometry, and the revolution's rate.*

### *Overall effects.*

Overall effects of a Ring Wing incorporation into Marine Vehicle architecture as shown for example on Fig. 2 are :
- Non-uniformity of the propulsor's inflow decrease.
- Noise and vibration decrease.
- Propulsor efficiency increase.

For more details on particular types of Vehicles – see below.

### *One shaft submarine.*

As you can see from test results for one shaft submarine with RFID (Fig. 3, 5) the flow-improving effect on the models is stable, the $\Delta$ may be lowered by as much as 60% and the discrete sound - by 6~10 db. The flow-improving effect expands to the state of maneuvering which is unique: no more devices are known that could automatically decrease the non-uniformity of the flow produced by the hull itself. Due to noise and vibration decrease while maneuvering, the upper limit of stealth speed of the maneuvering vehicle can be set higher.

The tests show as well that in case of small clearance between a hull and the RFID, the diffuser effect is observed, and the flow may be slowed down considerably due to pre-separation conditions. It is possible to use that effect for the improvement of the flow, but this may cost you some noticeable drag resistance increase.

### *Two shafts submarine and displacement ships.*

In this case, it is not rational to use RFID to suppress the shear general non-uniformity $\Delta$. The more delicate work is needed here which requires in full the implementation of the iterative design process described below in section G. The RFID design results shown in Fig. 6 have been received within a reasonable number of design iterations, but more important is that without the proposed iterative design method a tangible decrease of discrete sound could have hardly been achieved at all ( see the explanation in section G).

### *Other types of Marine Vehicles.*

On torpedoes and torpedo-like Vehicles flow improvement as a result of ring stabilizer implementation is the strongest (Fig 7,8 ). The general non-uniformity of the flow $\Delta$ can be practically eliminated in the case of rectilinear motion. Besides, there is a noticeable flow-improving and noise-decreasing effect on a maneuver. On small submarines and submersibles there may be several advantages received with a Ring Wing implemented: stabilizers span decrease, maneuverability and stabilization improvement, flow improvement with vibration and noise decrease, and propeller construction simplification.

## D. Relative risk

*Provided the described in section G complex of physical testing and computations had been performed and resulted in a noticeable numerically expressed prognosis of improving effect, then:*
*For each particular project, the positive effect is estimated in propulsors efficiency increase, noise and vibration reduction (in dB), vehicle speed increase, etc. compared to the case of conventional design or design without a flow-improving device. The estimations should be based on physical and mathematical model testing and computations (for instance - noise and vibration computations ). The maximal risk may be estimated as the cost of the flow-improving device removal and a propulsor replacement.*

### *For different types of Marine Vehicles:*

### *One Shaft Submarines.*

All types of Ring Wing implementation can be considered to have a high probability of success.
The estimated effect is high (see previous section), and the risk of over-estimation of the improving effect is low (provided all the tests and computations are done).

### *Two Shafts Submarines and Displacement Ships.*

Recommended types of arrangements with a ring wing: RFID, asymmetrical pre-ducts, and full ducts (see the last section of this document). The risk of over-estimation of the flow-improving effect is higher than for a single-shaft submarine. Full-scale experiments may be needed to confirm the results of model testing, computational modeling, and iterative design (described in section G of this document).

### *Small Submarines, Submersibles, Torpedoes.*

Here we have the firmest effect and a low risk of overestimating a positive result (see previous section).

## E. Efficiency and cost impact of new technology.

*The core of the proposed technology consists of the method of inflow harmonics control, the interactive design procedure based on this method (see section G), and the set of flow-improving devices and arrangements with a Ring Wing. The proposed interactive design procedure is expected to reduce the cost of the flow-improving and energy-saving devices because the procedure simplifies the task of inflow velocity distribution control. The noise and vibration optimization effect as a result of new technology implementation (design method and new devices ) is expected to be higher compared to current technology. The construction of Propulsors is expected to be simplified as a result of the proposed technology implementation: the wake inflow can be improved effectively for a propeller with a simple design and a low number of blades. The effect of sound and vibration decrease is expected to be stable and high. The energy-saving effect is expected in some cases. The proposed interactive design method is the only one currently known that **formally converges** to the task of acoustic signature decrease ( discrete spectrum sound decrease ). The method allows the formulation of simple and reachable goals of inflow harmonics control so it may be considered the first design method of such a kind providing a guaranteed result – design cost reduction could be derived from this fact. The details of a proposed technology with theoretical and experimental proof of its efficiency can be found in other sections of this document.*

## F. Signature control impacts.

*Signature control impact is expected to be the major area of improvement, this is reflected above – in section C.*

## G. Proposal for detailed technical design study with cost estimate.

*Design Procedure Description. Design iterations. Subtasks, time allotment, and cost estimation.*

### *Preliminary Calculations.*

For axis-symmetrical vessels here are the design steps that can be pre-calculated :
1. The optimization of ratios between the major dimensions of an RFID and the local diameter of the hull from the condition of maximum induction of vortex shroud when moving at an angle of attack. To solve that problem the hull is imitated with a dipole in the plane of an attached vortex of a ring wing which drives the problem down to the equation:

$$\partial A(r, b)/\partial r = 0,$$

   where:

$$A(r, b) = \frac{1-(r_0/r)^2}{1+\frac{2\lambda}{\pi}};$$

2. The optimization of the dimensions of RFID from the condition of smooth (non-separating) flow in the channel formed by a ring and a hull. The limit ratio between the *in* and *out* sections area is used here: $S_{out}/S_{in} = 1.2$, which ensures a non-separating flow inside a diffuser.

3. The estimation of a maximum opening angle of an RFID meeting the condition of non –separating flow on the under-pressured (sucking ) side of the wing. The ring is devised into segments with angular dimensions defined by maximums of the function $\partial \Gamma / \partial \varphi$. The empirical formula for the critical angle of attack may be used here: $\varepsilon_{cr} \approx 30°/\sqrt{\lambda_p}$ , where $\lambda_P$ - aspect ratio of a segment.

4. The estimation of the circular variation of the chord of an RFID from the condition of a maximum effect on circular distributions of the velocities on the given radius of a propeller's disk.

A simplified formula $b = \Delta_b \cos 4\varphi + b_0$ is accepted for a chord. The axis-symmetrical boundary layer is pre-calculated using any method from the shelf. Besides this, the following assumptions are enforced :

a) the cross-shaped symmetrical stabilizers are the sole assembly of parts jutting out of the hull,
b) the stabilizers affect the flow only with their head vortices,
c) the threads of head vortices of stabilizers remain in meridian planes laying under angles $\varphi = \pi(2k+1)/8$ to the vertical plane.

The value of $\Delta_b$ is found from the condition of zero shift of the intersection of the given flow line (laying in the plane of a stabilizer) with the plane of the propeller's disk (Fig.1 ).
These calculations allow to draw some diagrams to estimate the geometry of an RFID.
The final design of an RFID for a single-shaft submarine is to be done on the basis of model experiments with RFID.

As for two-shaft submarines and other non-axis - symmetrical vessels   - the series of experiments with stepwise adjustments of the construction is practically the only way to design RFID.

## *Iterative design.*

### *Theory of harmonics control.*

The main part of the RFID design is supposed to be done as a series of adjustments based on model test results. If Computational Hydrodynamics could provide a satisfactory method of flow calculation downstream of the RFID, the physical model tests would be substituted with computer model tests. In any case, it is necessary to make a series of iterations to find an optimal geometry of RFID. The same is true for any "flow-improving" modifications of the vehicle's architecture.

In many cases the substantial decrease of $\Delta$ - general non-uniformity of the flow (see section B and Fig. 3) can be achieved. There is no guarantee though that the particular harmonic $a_i$ contributing the most to the level of discrete sound would be decreased as a result of $\Delta$ decrease.

Looking at the circular distribution of a longitudinal velocity $V(\varphi)$ an engineer or operator always can define some *controllable* characteristics of this curve. For instance: the removal of a stabilizer would kill a certain peak on $V(\varphi)$, and the placement of a flat–wing vortex-generator on a stern part of a hull would produce a peak or wave on $V(\varphi)$ within some predictable sector $[\varphi_1,\varphi_2]$. Practically all the controllable characteristics of the circular velocity distributions $V(\varphi)$ at different radii of the propeller are the "rough" ones compared to the real aim of the control - the harmonics of $V(\varphi)$.

The $\Delta$ may be considered as the roughest one, all the others – as more detailed.
For instance :

- maximum derivative of $V(\varphi)$,

- number N of local extremums within the interval $[0,2\pi]$ ,

- full deviation $\mathbf{D}(V(\varphi)) = \int_0^{2\pi} |dV(\varphi)|$

The very control of $V(\varphi)$ is in a sense a process of putting certain not fully defined peaks on that curve within some sectors.  To choose optimal sectors for the peaks and to predict their effect on the harmonics of the curve $V(\varphi)$ the theory is required.
Such a theory has been created based on the following :

a) theorems of **Fourier Coefficients estimations** that deliver **relations** between **harmonics** and **rough characteristics** of a curve:

$$|a_i| < \frac{2\Delta}{\pi}; \qquad |a_i| \leq \frac{D(V(\varphi))}{\pi i} ;$$

b) the concept of curves $V(\varphi)$ *skeletons* providing upper estimations of the modules of Fourier Coefficients (based on the above-shown formulae for Fourier Coefficients estimation).

c) features of **symmetry** of Fourier Coefficients,

d) the concept of an  *ideal curve* (the curve $V_{ideal}(\varphi)$ that does not contain harmonics whose suppression is a goal of the control),

e) the concept of  a *limiting $\delta$ -wide band* (the area between two curves defined as $V_{ideal}(\varphi) + \delta/2$ and $V_{ideal}(\varphi) - \delta/2$ ),  such that the fall of a certain harmonic amplitude in a certain predefined interval of values is guaranteed if the curve $V(\varphi)$ will remain within the band  along the whole interval $[0, 2\pi]$.

f) The concept of a *fuzzy peak* – a subset of a band described above.

Based on this theory *reachable* goals of V($\varphi$) control (goals of iterative design of RFID) can be formulated as follows: to put V($\varphi$) within the predefined band at the predefined ideal curve $V_{ideal}(\varphi)$.

The Ideal Curves $V_{ideal}(\varphi)$ by definition are the ones in which particular target harmonics are equal to 0.
Theoretically, there is an infinite set of such curves for any particular set of target harmonics.
For our goals, the Ideal Curves may be subdivided into 4 basic types (Fig 9)
- Trivial Ideal Curve $V_{ideal}(\varphi)$ =const (horizontal line on the circumferential diagram).
- Dynamic Ideal Curve – the one equal to a sum of a certain small number of major harmonics of the initial V($\varphi$). This one is supposed to catch a base pattern of the initial V($\varphi$).
- Optimal Ideal Curve—the one received from the initial by subtracting only the target harmonics. This one is geometrically the closest; it delivers the minimum mean square deviation by definition.
- Partial Ideal Curve – generalizes a powerful set of ideal curves with a following pattern: they only deviate from the initial one within particular sectors of the interval [0,2$\pi$] and coincide with the initial one anywhere outside these sectors.   Peaks are the subset of this Ideal Curve type.

Ideal curves for each design case should be chosen according to the base pattern of the inflow V($\varphi$) implied by the base architectural pattern of a vehicle. For instance, the curves V($\varphi$) of two – shaft submarine or one-shaft displacement ship have a base pattern similar to sin($\varphi-\varphi_0$), so
- Dynamic Ideal Curve would be a rational choice for a two-shaft submarine and one-shaft displacement ship.
- Trivial Ideal Curve is acceptable for one shaft submarine since the ideal architecture in this case is a pure rotation body which implies V($\varphi$) = const.
- Optimal Ideal Curve acceptance for design iterations implies that the inflow can be affected at any particular sector. This is a case of RFID design.
- Partial Ideal Curve and its subset – fuzzy peak should be used when some sectors of the inflow are not accessible.

Expected *number of local extremums* **N**  (Fig. 9) of V($\varphi$) is an important parameter for limiting band width calculation. It is used as well for upper estimation of skeletons' harmonics.
Details of the described theory have been published in [ 11].
The following issues are addressed there as well:
- relations between V($\varphi$) *systematic error* and *harmonics calculation systematic errors*
- harmonics decrease *probability* in cases when V($\varphi$) fits into a band *wider* than predefined $\delta$.

Fig. 10 shows the first steps of an iterative design based on the described theory. This is a stabilizer design for a small submarine. Harmonics change prediction has been done differently at each step: at the first step, based on skeletons, and at the second step, based on a limiting band in the proximity of a Trivial Ideal Curve.

## *Detailed description of the design procedure.*

## *Case of pure RFID.*

This is a case when the Ring Wing works as a pure flow-improving device affecting the flow with its longitudinal vortices and having low side effect on stabilization.  This case is shown in Fig.2 (arrangements 1  and 6 for single-shaft submarines)  and Fig 6 - for two-shafted submarines.
 and displacement ships.

Here below are the subtasks and steps of the design. ***The flowchart*** of the design procedure follows the listing of steps.

1. Dimensions estimation using the formulae provided for the case (see above ) or using the diagrams based on these formulae.

2. The sketch design of struts: struts number, dimension, position, and way of fixing on the hull or appendages.
3. Choice of a Ring Wing profile.
4. Estimation of stabilization effect (for the case of single–shafted vessel – the formulae and diagrams on cross–ring wings integral comparison can be used - Fig 4).
5. Adjustment (decrease ) of the area of initial stabilizers.
6. Ring model technical design (based on the construction schemes above) and drawing.
7. Vessel's model design and manufacturing ( see the appropriate cost in David Tailor, Krylov, and other experimental Basins)
8. Ring model manufacturing.
9. The initial model test: measurement of the velocity distribution in the disk of the propeller without the model of the flow-improving device.
10. Fourier Analysis of the initial curves.
11. Initial computation of discrete spectrum sound for several models of a propeller in the absence of Flow Improving Devices. Requisites: discrete sound computation software, computer time.

*This step can be deferred in some cases when the Fourier Analysis of the initial curve made at a previous step shows a relatively high amplitude of target harmonics. In such a case all sound computations could be done at the very end of the design procedure. The knowledge of the details of the Filter Action of the propeller allows us to go to the next step without waiting for the results of this computation.*

12. Computation and build of ideal curves, skeletons, and bands (based on the proposed formulae and algorithms –see previous subsection ).
13. Computation of discrete sound for an ideal curve and skeleton curve composed of segments of the initial curve and bands.
14. Iterative process of (inflow measurement + comparison to the pre-calculated bands ) -> ring shape adjustment -> (again inflow measurement + comparison to the pre-calculated bands ).
15. Final calculations of a propeller discrete sound.
16. In case of good computational prediction -  constructive adjustment of ring, struts, appendages and /or stabilizers geometry, technical drawing.
17. Final model test and adjustments (may be omitted if the final design is not different from the one tested on a model)

*To facilitate the iterative design procedure, it is convenient to make models of an RFID with narrow changeable rear flaps. The final shape of the rear edge of RFID is supposed to be drawn smoothly.*

18. Full-scaled RFID design and drawing.
19. Manufacturing of a RFID.
20. Montage on a real vessel (time and costs to be defined for a particular project)
21. Full-scale tests with sound and vibration measurements (on the first ship of a series).

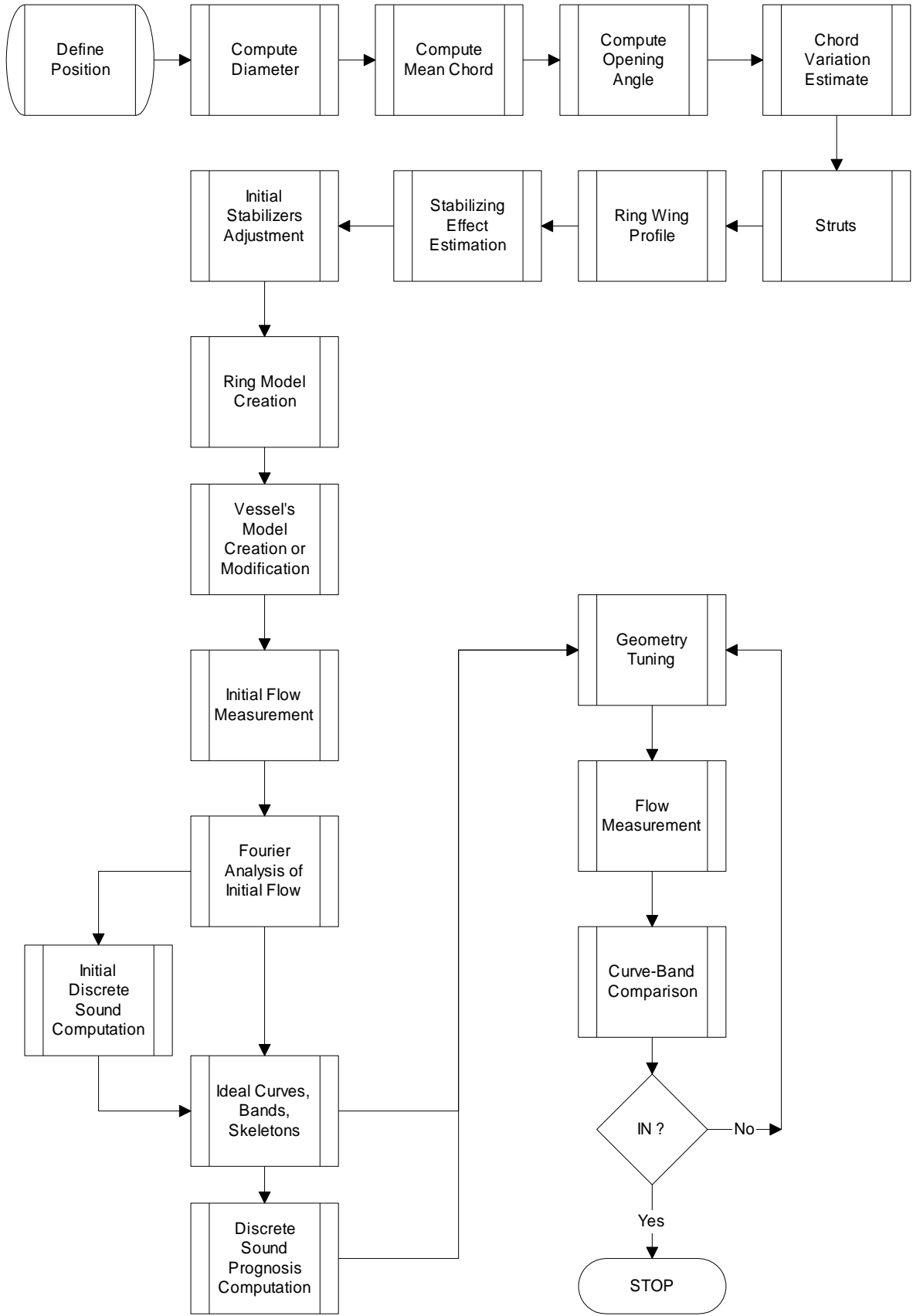

## *Case of Ring Wing as stabilizer and/or rudder*

These are the cases applicable to underwater vessels when the flow-improving effect comes essentially from the substitution of conventional cross-shaped stabilizers with Ring-shaped one. These cases correspond to arrangements 1, 3, and 5 of Fig .2.
Here below are the steps of such a design.

1. Pre-calculation of a ring stabilizer's dimensions. The comparative diagrams of ring wing versus cross-shaped wing dynamics and geometry ( Fig .4 ) can be used here, as well as the corresponding formulae.

2. The design of a Ring Wing–based rudder-stabilizer system as a whole.
    - rudders mechanical design (see arrangements 1 , 3, and 5 of Fig.2 )
    - rudders hydrodynamics design
    - ring wing struts dimension calculation
    - struts effect on maneuverability and stabilization calculation
    - ring wing dimensions adjustment with account of struts

    Here conceptual mechanical design is needed ( especially for arrangements 3 and 5 of Fig.2 ).

3. Stabilizers models drawing and manufacturing. Two arrangements at least must be manufactured for the comparison – conventional (cross-shaped) and one of the new ones  (of 1,3,5 arrangements from Fig.2).

4. Vessel model design and manufacturing.

5. Initial model comparative tests with at least two arrangements of stabilizers.
    - inflow measurements
    - drug resistance measurement
    - standard stabilization and maneuverability characteristics test

6. Test results comparison and estimation of acceptable drug increase in trade of sound decrease. Decision making on the continuation of design: choice between conventional design and the one with a Ring stabilizer.   If the conventional design proves to be better – stop, otherwise – go to the next step.

7. Computation of discrete spectrum sound for several models of a propeller in the flow behind the Ring stabilizer.

8. Optional step: Ring Wing stabilizer geometry, struts shape, and position iterative tuning for discrete sound optimization. The proposed method of inflow harmonics control should be used here  (see positions 11 and 12 of the previous section).

## *Unconventional propulsors design.*

The Ring  Wing has been proposed as an effective flow-improving device for a propeller.  Thus, its design always refers to the one of a propeller and vice versa: the propeller diameter and blades number are needed as an Input for the Ring Wing design, and the flow field behind the Ring is the Input for a Propeller design. Beyond this, the two are clearly separate for most of the arrangements with a Ring Wing shown so far in this document.  When the Ring Wing is placed far from a  Propeller its effect on a propeller is strong, while the effect of a propeller on the Ring Wing via the pressure redistribution is negligible so the assembly of two cannot be considered as an unconventional  Integrated Propulsor (to clarify the term - see [7]). Sometimes though, a Ring Wing as flow improving device may need to be placed close to a propeller due to general design restrictions. Then in cases when the duct is needed for propulsor efficiency, it could be reasonable to adjust the duct's geometry to decrease noise and vibration.

Thus, assembling a non-axis-symmetrical Ring Wing and a Propeller results in an Integrated Propulsor.

Two different arrangements of such a propulsor with a non-axis-symmetrical Ring Wing may be considered: the propeller with a pre-duct and the completely ducted propeller.

The design of such an assembly is a complicated task. The design should be organized as an iterative process with computations and model tests. At the first stage of the design (which are parametric studies and preliminary computation) the Ring Wing may be considered axis symmetric. This design stage could be effectively done using the design method proposed for Integrated Propulsor by Kerwin et al [7]. At the second stage of the design, the details of a non-axis-symmetrical pre-duct/duct are supposed to be elaborated. This should be done with extensive model tests as a sequence of iterations as it was described for the case of RFID. The struts' position and shape will be very important here. Struts number and position proved to have a strong effect on the flow details which define the level of propeller noise (see the design iterations for the case of ring stabilizer for a small submarine in Fig 10. ). The proposed method of inflow harmonics control with predefined bands and skeletons for inflow velocity circumferential distribution allows using of the struts as an effective means of such control. Struts arrangement could be used as well as a pre-swirl. The main complication of the iterative design of a non-axis-symmetrical duct/pre-duct versus a design of the non-axis-symmetrical Ring Wing placed relatively far forward of the propeller is that in case of a duct or pre-duct, the Propeller inflow (as the Input for Propeller sound and vibration calculation ) should be measured as an effective one i.e. - in presence of a working propeller. This propeller model though should not be modified at each iteration of the design when the duct or strut geometry is adjusted. The very axis asymmetry of a duct – the main agent of flow details control - is supposed to be a circumferential variation of the chord – similar to how it has been proposed for the Ring Wing placed far forward of the propeller.

The principles of the work of such a non-axis-symmetrical pre-duct or duct within an Integrated propulsor are different from the one of a RFID described above. The effect on the propeller inflow is now not a result of the work of passively generated longitudinal vortices in the boundary layer of a hull, but a result of direct interference between pressure fields of a Ring and the blades of a Propeller.

As for the task of discrete spectrum sound and vibration optimization, the iterative design procedure based on predefined bands for the propeller's inflow velocity distribution described above is entirely applicable here. For model tests, the models of non-axis-symmetrical ducts can have the same construction with changeable rear flaps as the one for RFID described in the previous section.

**References.**

9. S. Cordier, F. Legrand, J.C.Pinard "Hull and Shaft Wake Interaction" SNAME Propellers/Shafting 97 Symposium, 1997.
10. Larsson L., Broberg L., Keun-Jae Kim, Dao-Hua ZhangA Method for Resistance and Flow Prediction in Ship Design. SNAME Transactions, vol.98, 1990, pp.495-535.
11. Sluchak V., "Method of control of velocity distribution harmonics forward a Propeller", The Shipbuilding Industry, RUMB, ISS, 6s, 1986, Leningrad. (in Russian)
12. Sluchak V., "The thin ring wing as a means of flow improvement upstream of a propeller", SNAME Propeller/Shafting'97 Symposium, Virginia Beach, VA., September 1997

**Figures.**

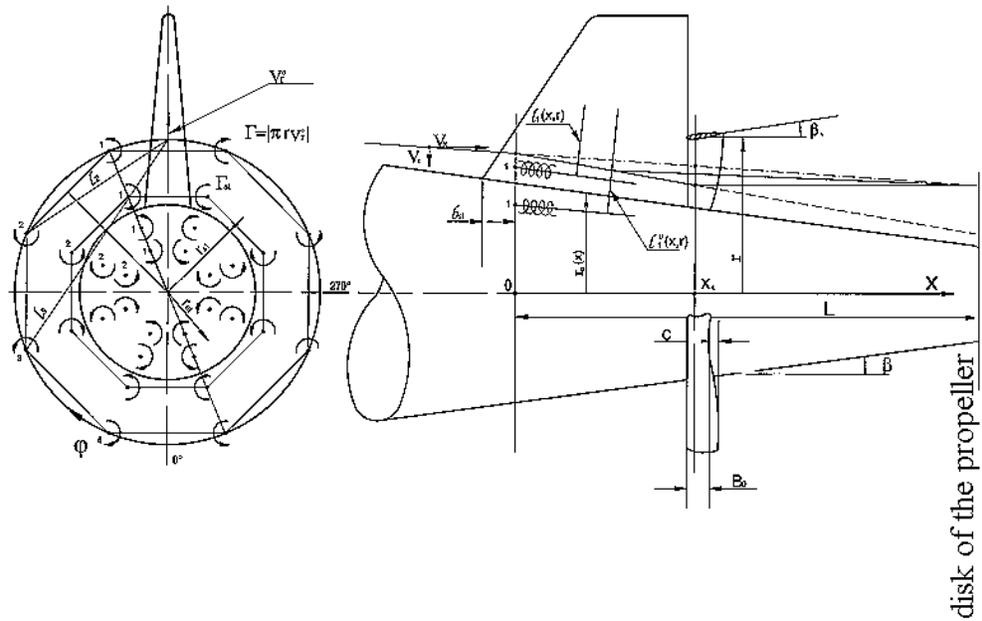

**Fig 1** . Vortex model of the flow around stabilizers and RFID

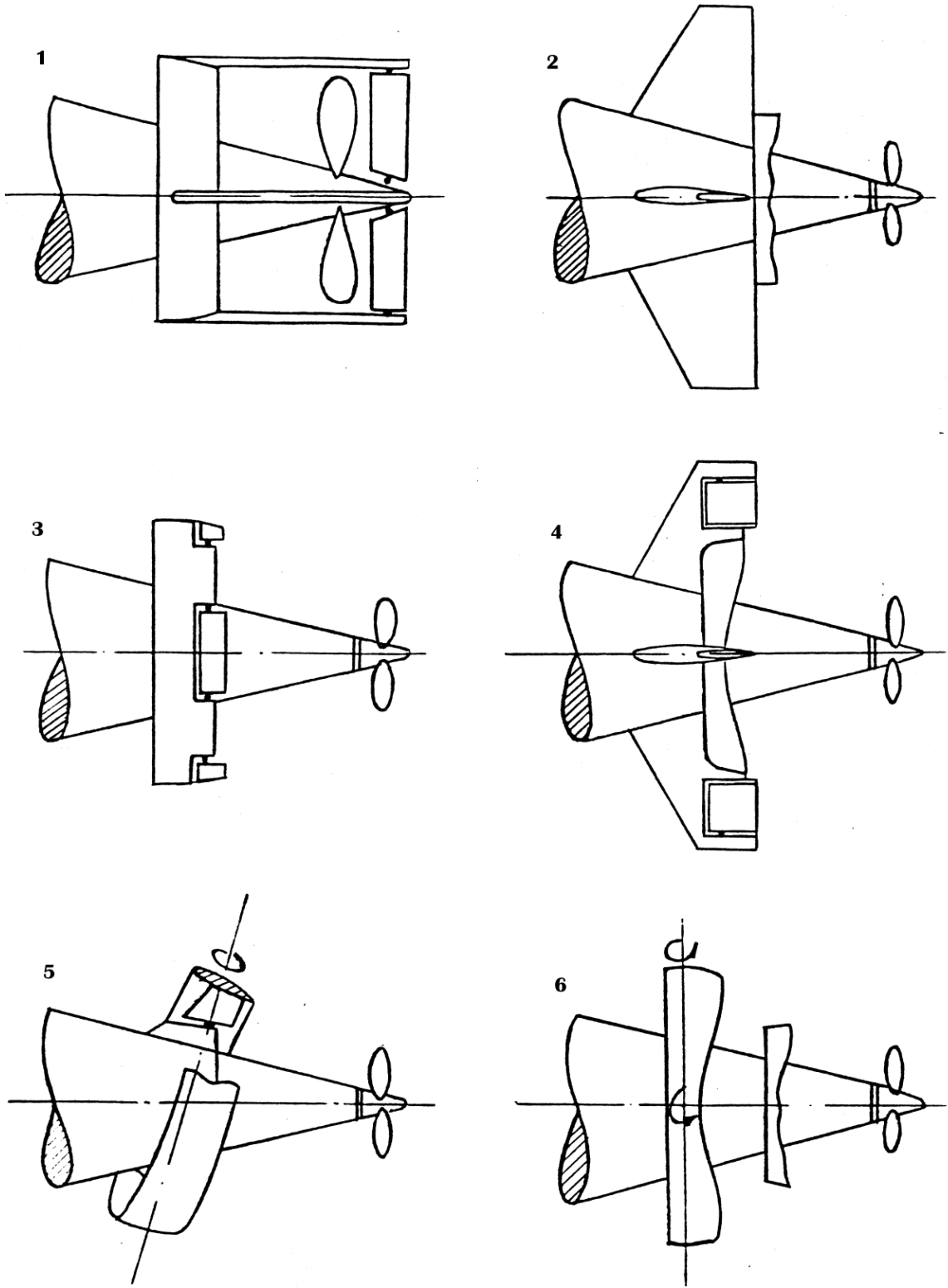

**Fig 2**. Stern arrangements with a ring wing as a stabilizer, rudder, and flow controller (RFID).

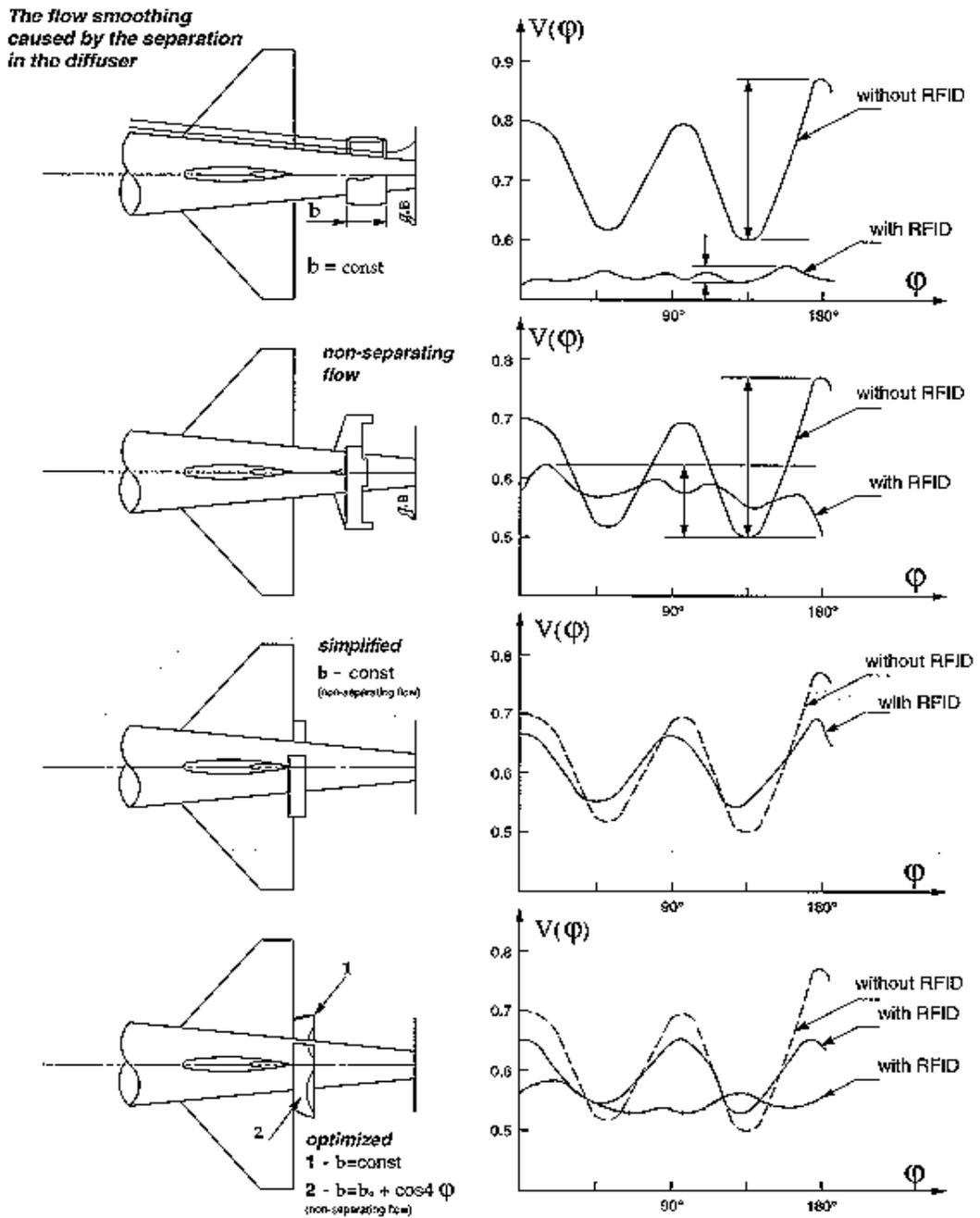

**Fig 3** Stability of RFID effect.

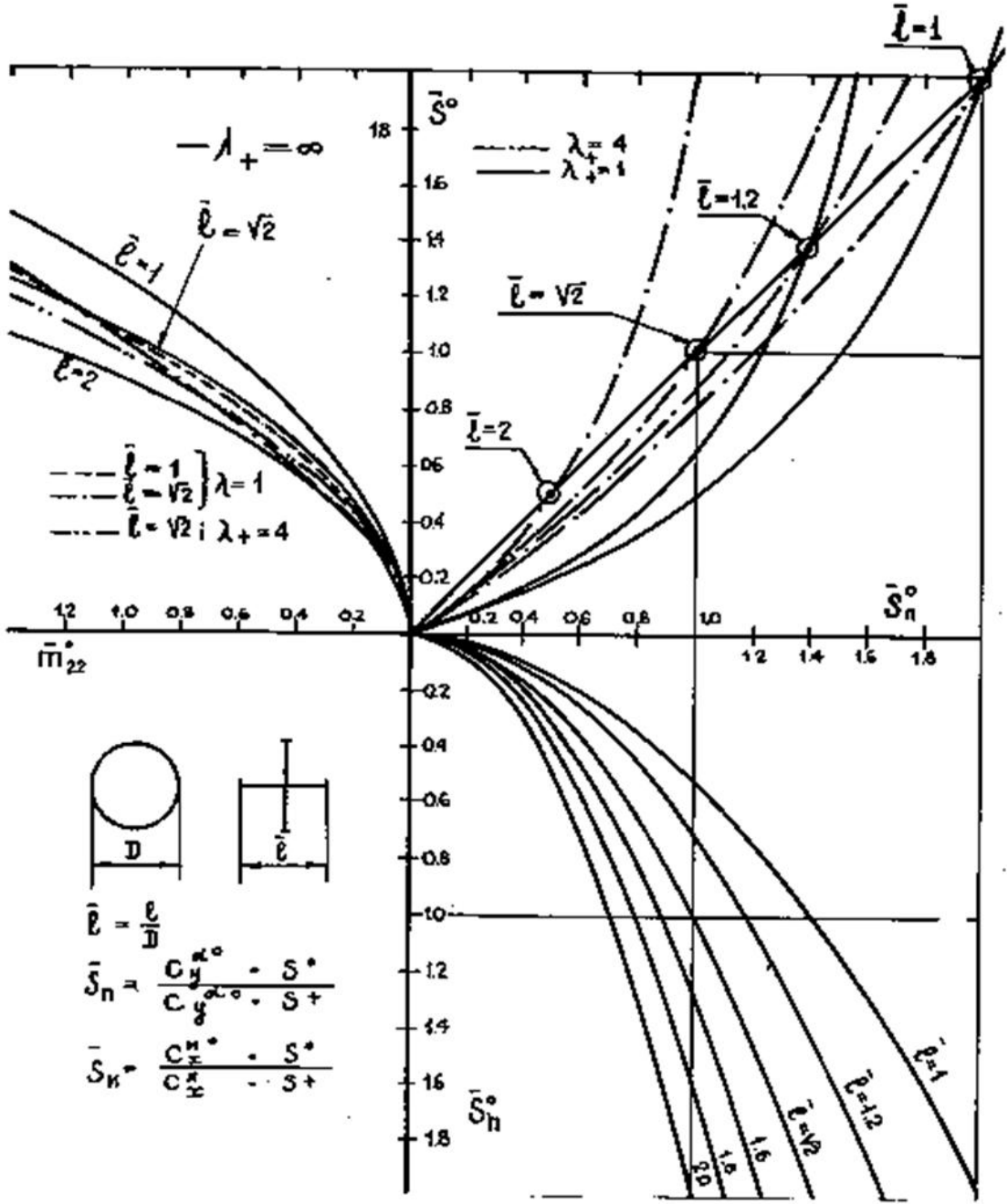

**Fig 4** Integrated comparison of Ring-Shaped [O] and Closs-shaped [+] wings.

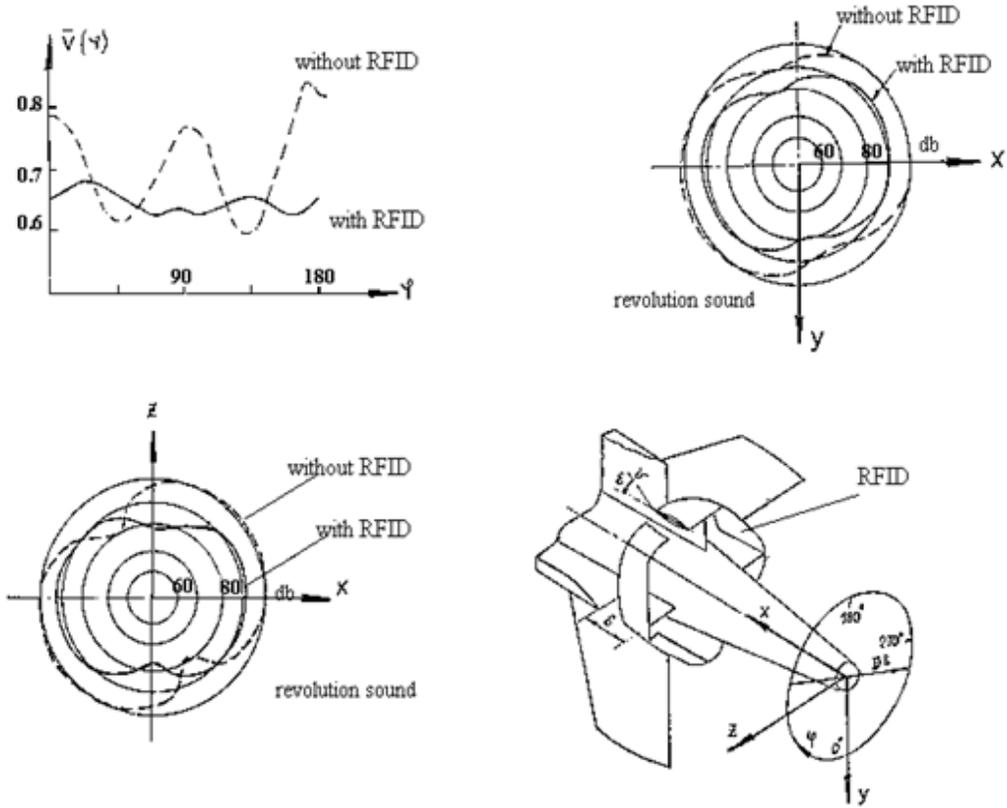

**Fig 5** Effect of RFID on single shaft submarine (model tests and computation).

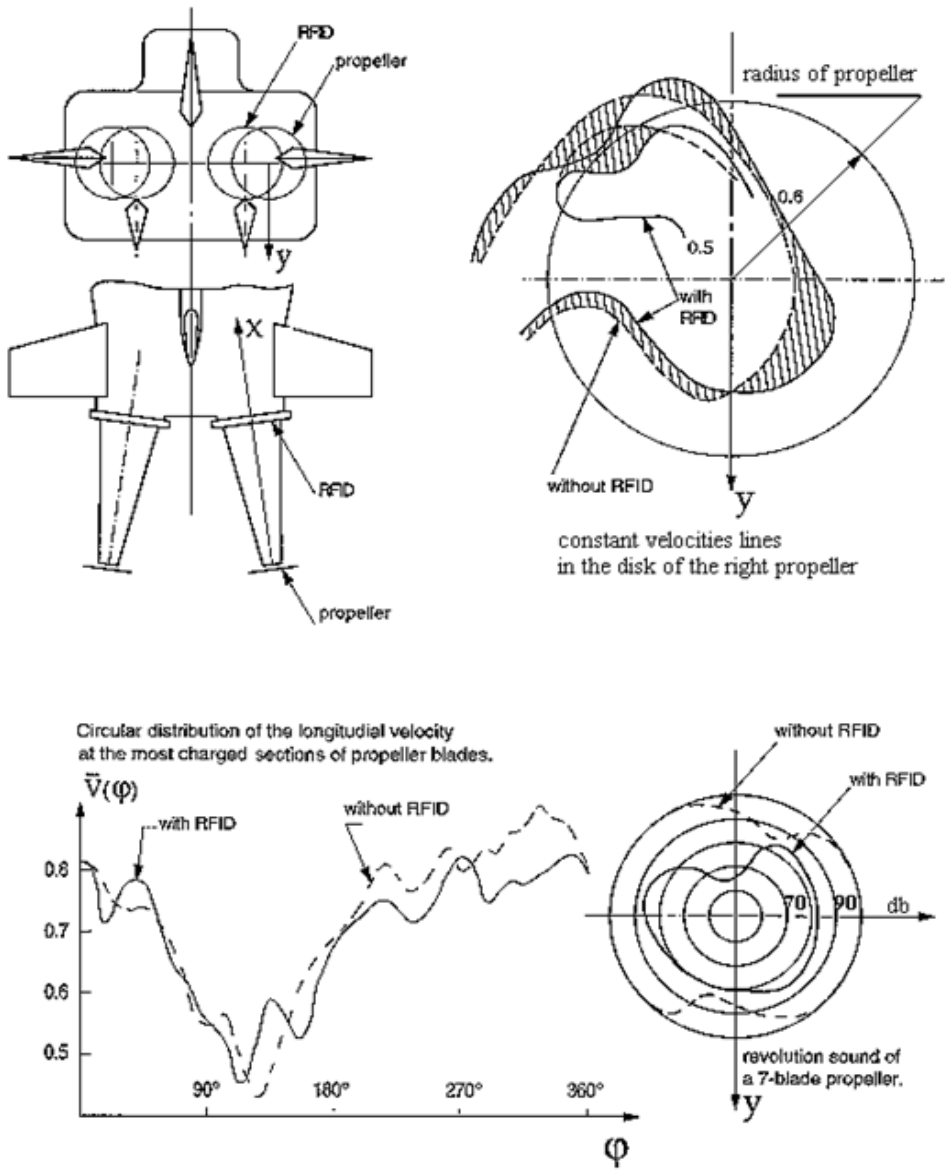

**Fig 6** Effect of RFID on a two-shaft submarine (model tests and computation).

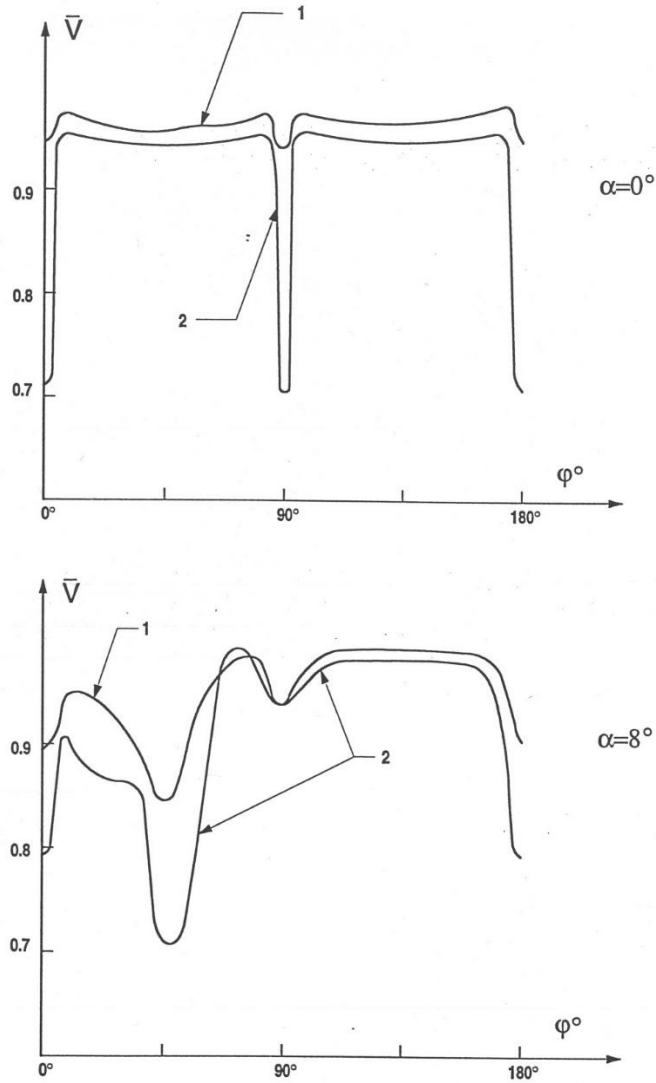

**Fig 7** The results of cross-shaped stabilizer substitution by a ring wing on a torpedo. 1 – ring wing, 2 – conventional cross-shaped wing.

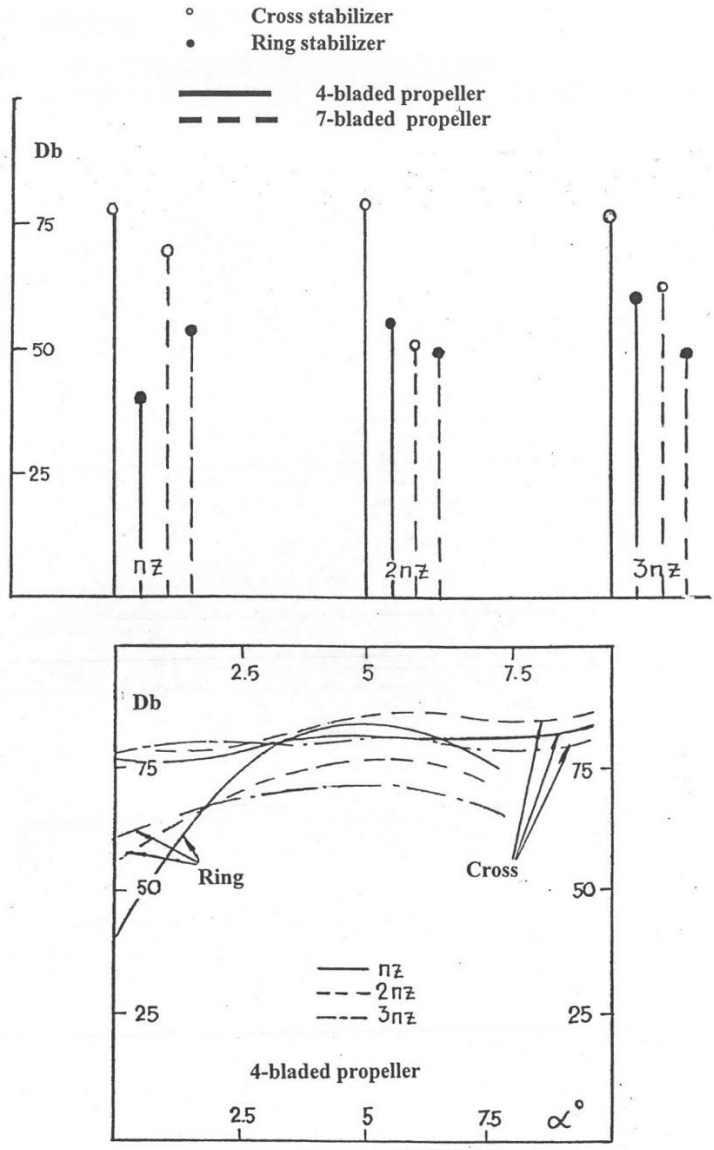

**Fig 8** Discrete spectrum sound calculation for a torpedo with different propellers (4 and 7) blades, and stabilizers (cross and ring).

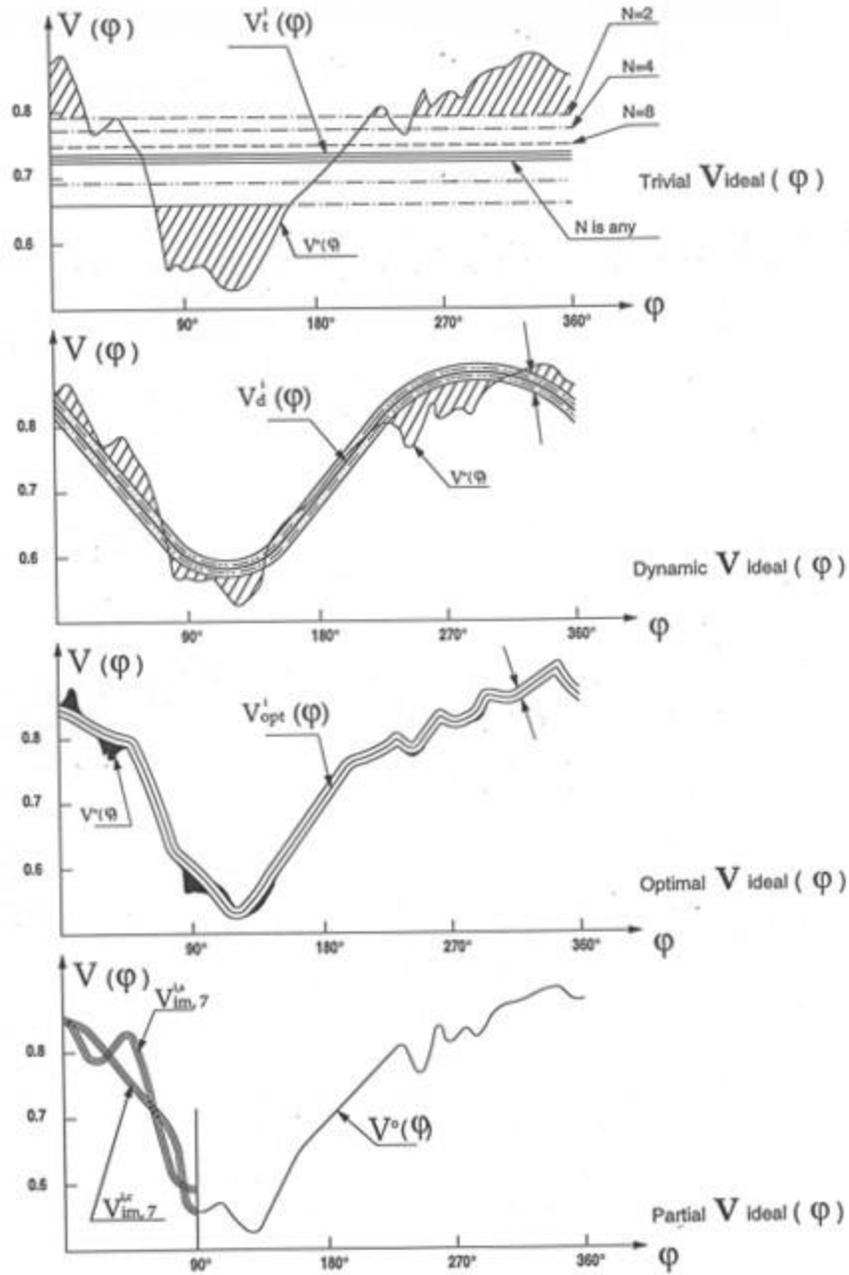

**Fig 9** Ideal curves and limiting bands.

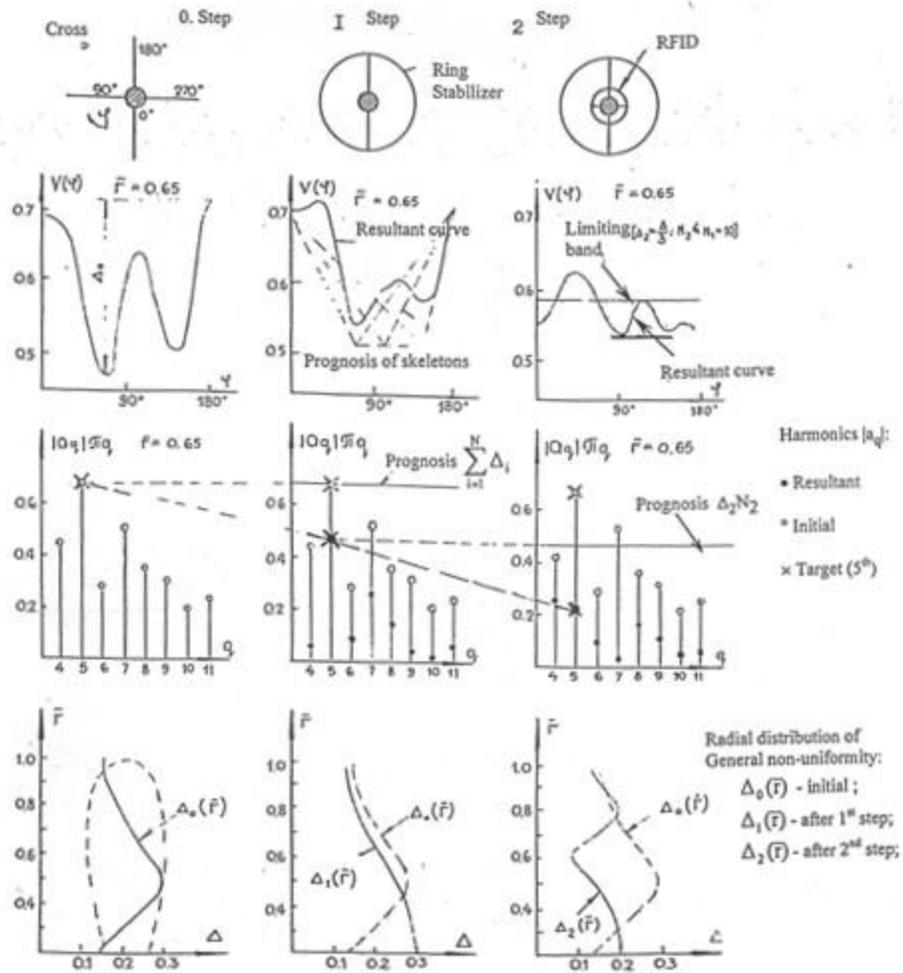

**Fig 10** Two first steps of iterative design of the submarine stabilizers with a target to decrease the 5[th] harmonics of circumferential distribution of inflow velocity at relative radius =0.65. At the design's first step, the initial cross stabilizer is substituted with a ring mounted on two struts. In the 2[nd] step, the Ring flow-improving device is mounted between the propeller and ring stabilizer.